\documentclass[]{tPHM2e}

\begin{document}
\doi{10.1080/14786435.20xx.xxxxxx}
\issn{1478-6443}
\issnp{1478-6435}
\jvol{00} \jnum{00} \jyear{2010} 

\markboth{Taylor \& Francis and I.T. Consultant}{Philosophical Magazine}

\title{Screw symmetry in columnar crystals}

\author
{
A. Mughal$^{\rm a}$ $^{\ast}$\thanks{$^\ast$Corresponding author. Email: adil.m.mughal@gmail.com\vspace{6pt}}  
\\\vspace{6pt} $^{\rm a}
${\em{Institute of Mathematics and Physics, Aberystwyth University, Penglais, Aberystwyth, Ceredigion, Wales, SY23 3BZ, United Kingdom}}
 \received{v4.5 released May 2010} }

\maketitle

\begin{abstract}
We show that the optimal packing of hard spheres in an infinitely long cylinder yields structures characterised by a screw symmetry. Each packing can be assembled by stacking a basic unit cell {\it ad infinitum} along the length of the cylinder with each subsequent unit cell rotated by the same twist angle with respect to the previous one. In this paper we quantitatively describe the nature of this screw operation for all such packings in the range $1\leq D/d \leq 2.715$ and also briefly discuss their helicity.

\begin{keywords}Hard sphere packing, screw symmetry, phyllotaxis 
\end{keywords}\bigskip

\end{abstract}

\section{Introduction}

That nature creates forms and structures of great diversity according to the requirements of simple physical laws is a subject of endless fascination. The possible ways in which atoms, spheres or cells fit together into alternative arrangements depends on both symmetry and the nature of the physical forces involved. While these physical interactions maybe simple, nevertheless the high pressures encountered in strongly confined systems can compel molecules to adopt complex yet ordered arrangements. In such systems there exists an intimate connection between molecular morphology and the precise shape of the container. 

An example of this is the densest packing arrangement of equal sized hard spheres of diameter $d$, in an unbounded cylinder, of diameter $D$. Recently we have conducted numerical simulations \cite{mughal1, mughal2} (based on simulated annealing) which have so far identified $40$ distinct spiral structures in the range $1\leq D/d \leq 2.873$. Our simulations greatly extend the list of packings discovered by the earlier work of Pickett et. al \cite{pickett} and include both monolayer arrangements, for $1\leq D/d \leq 2.715$, (in which all the spheres are in contact with the confining cylinder) and more complex multilayer packings, for $ D/d > 2.715$, that include internal spheres (i.e. spheres which are in contact with other spheres but not in contact with the cylinder). 

We have dubbed \cite{mughal2} all such quasi-one-dimensional packings as {\it columnar crystals} since they are periodic; each structure can be assembled by stacking unit cells {\it ad infinitum} along the length of the cylinder with each subsequent unit cell rotated by the same twist angle with respect to the previous one. In this paper we detail the nature of this screw operation for all monolayer packings and the implications this has in terms of helicity. 

Further details, including the volume fraction (density), chirality and the number of spheres in the unit cell, for the packings, can be found in \cite{mughal1, mughal2}. Some of the simpler packings that we have tabulated have been realised in experiments with dry foams \cite{pittet1, weaire, pittet2, boltenhagen, hutzler} and more recently using wet foams \cite{meagher}. Similar, monolayer arrangements have also been observed in a biological microstructure \cite{erickson}, by confining colloids in micro channels \cite{moon1, moon2, badel, tymczenko} and encapsulating fullerenes within nanotubes \cite{Khlobystov, Yamazaki, Warner}.

\begin{figure}
\begin{center}
\resizebox*{10cm}{!}{\includegraphics{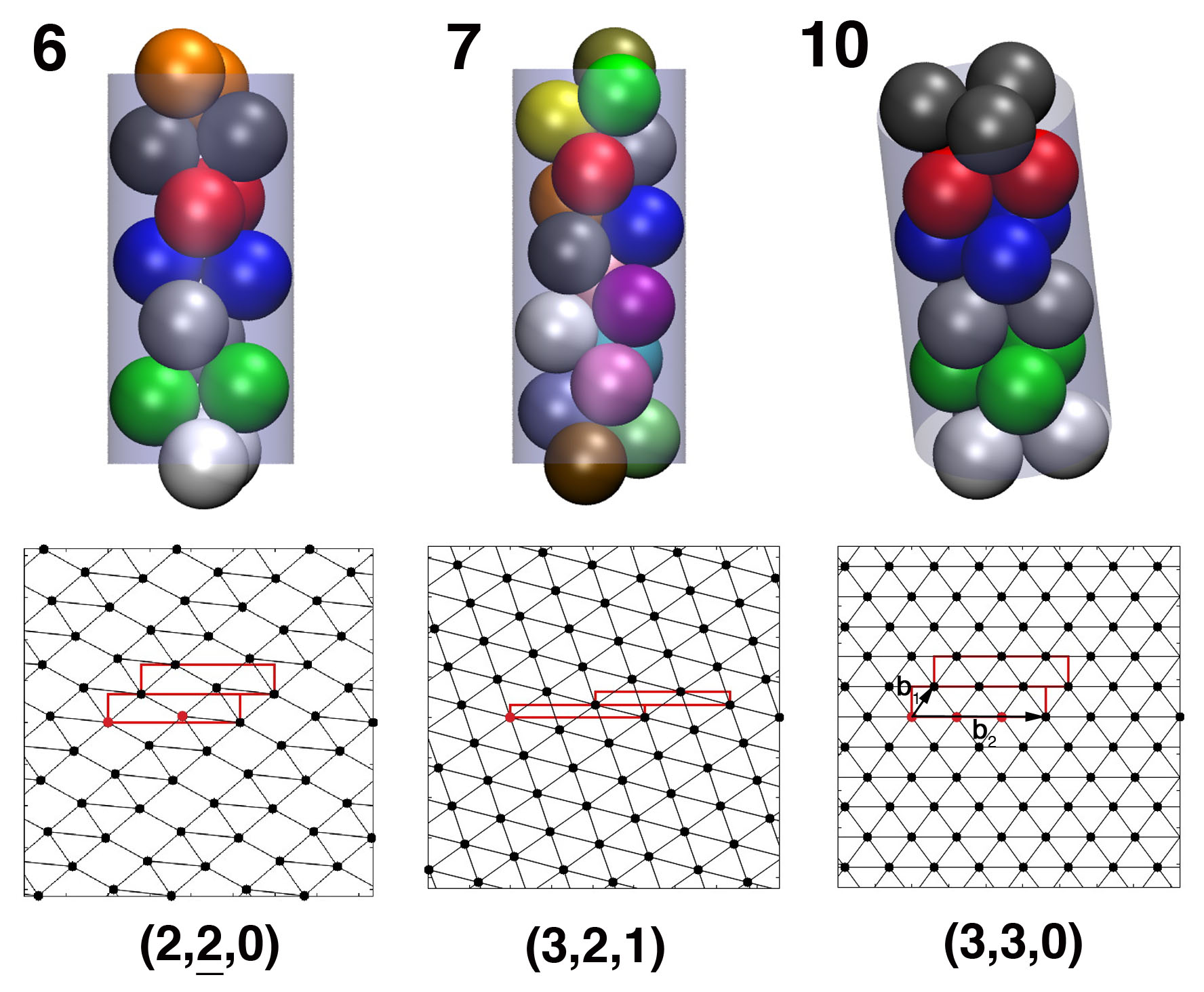}}
\end{center}
\caption{{\bf Left}: Packing 6 is a line-slip structure found in the range $2.0 < D/d \leq 2.039$. The unit cell contains two spheres. Note not all spheres in packing $6$ have six contacts, varying the diameter ratio $D/d$ changes the relative distance between next nearest neighbours until at $D/d =2.039$ there are 6 contacts per sphere giving rise to packing 7. {\bf Middle}: Packing 7 is a chiral maximal contact packing found at $D/d =2.039$, it contains one sphere in the unit cell. {\bf Right}: packing 10 is a maximal contact achiral packing which contains three spheres per unit cell. For illustrative purposes the basis vectors ${\bf b}_1$ and ${\bf b}_2$ (see text) are shown in the rolled out diagram in this final example. For each packing the corresponding phyllotactic diagram is shown below and assigned a unique phyllotactic index \cite{mughal2}. The phyllotactic  pattern can be generated from a basic tile shown by a red rectangle}
\label{packing_images}
\end{figure}

Our numerical method makes use of novel twisted periodic boundary conditions, whereby the densest packing is sought for a unit cell containing $N$ spheres by systematically varying both the length of the simulation cell and the twist angle. Full details of this protocol have been reported previously \cite{mughal2} and the nature of the screw periodicity is briefly summarised below. The validity of this approach has been confirmed independently by Chan \cite{chan} using numerical sequential deposition, which in contrast does not employ twisted periodic boundary conditions. Both methods (deposition and annealing) are in agreement as to which arrangements are the densest for a given value of $D/d$ for all monolayer packings (i.e. in the range $1\leq D/d \leq 2.715$).

In the case of monolayer packings we have developed analytical approximations which elucidate the nature of these structures \cite{mughal1, mughal2}. Our approach is based on the method of {\it phyllotaxis}, which originates in biology but is now used for the classification of triangular patterns on the surface of a cylinder in any context \cite{erickson}.

Examples of some single-layer packings are shown Figure~\ref{packing_images} (see  \cite{mughal2} for the full sequence of packings), along with the corresponding ``rolled out''  phyllotactic pattern for each structure. Such patterns are generated by mapping the coordinates of the sphere centres $[D',\theta,z]$ (where in a monolayer packing, since all the spheres touch the cylindrical boundary, all the sphere centres lie on an inner surface of diameter $D'=D-d$) onto the plane using Cartesian coordinates $[(D'\theta/2), z]$ where they are shown as black dots. Contacting spheres are indicated by dots connected by black lines. 

Monolayer packings are found to be of two types: maximal contact or line-slip structures. Maximal contact structures (such as packings 7 and 10, see Figure~\ref{packing_images}) are those in which each sphere touches six neighbouring spheres. These packings exist only at discrete values of $D/d$ and can be classified using a set of three integers known as phyllotactic indices $[l=m+n,m,n]$. Line-slip packings, however, involve a relative slip between neighbouring spiral chains which results in fewer contacts on average per sphere (as can be seen in the rolled out diagram for packing 6). As a consequence of this degree of freedom, line-slip structures exist over a range of values of $D/d$. The direction of the line-slip is indicated in the phyllotactic notation by the use of bold (or underlined) numerals \cite{mughal2}. 
 
\section{Screw symmetry}

Our simulated annealing search is confined to structures that are periodic in the following sense. There is a primitive cell, of length $L$, containing $N$ spheres, the structure being generated from this by the screw operation of (i) translation along the cylinder axis by $nL$ (where $n$ is any integer) combined with (ii) rotation about the axis by an angle $n\alpha$. This screw operation represents the underlying symmetry of columnar crystals \cite{mughal2}. 

Similarly, the corresponding phyllotactic diagram for a given packing can be reduced to a basic rectangular tile, which contains a number of sphere centres (coloured red in Figure~\ref{packing_images}). This tile can then tesselate the plane using the basis vectors ${\bf b}_1=[\alpha D'/2, L]$ and ${\bf b}_2=[(2\pi) D'/2,0]$, as for example indicated in the rolled out diagram for structure 10 in Figure~\ref{packing_images}.

Both $L$ and $\alpha$ are plotted as a function of $D/d$ for all monlayer packings in Figure~\ref{screw_symm} (top and middle, respectively). Also shown is the twist angle per unit length $\alpha/L$ (bottom). The vertical lines indicate discontinuities in the derivative of the plotted quantities. For a particular value of $D/d$ our simulations produce left and right handed chiral structures with equal probability. Thus in Figure~\ref{screw_symm} we plot the values of the twist angle for only one of the enantiomers, the twist angle for the other enantiomer is then given by $\alpha'=2\pi-\alpha$.

We now describe Figure~\ref{screw_symm} in detail and illuminate this discussion by alluding to the structures shown in Figure~\ref{packing_images}. 

\subsection{Maximal contact packings}

Maximal contact packings are highly symmetric structures that are found at discrete values of $D/d$. All single-layer packings of this type are homogenous in the sense that all spheres are equivalent; each sphere is in contact with six neighbouring spheres (or five in the case the structure at $D/d=2$). These packings correspond to local peaks in the volume fraction (or packing density).

For particular values of $D/d$ the the optimal (maximal contact) packing structure can be easily surmised \cite{mughal2}. These structures are directly related to the analogous two-dimensional problem of finding the smallest diameter circle into which N non-overlapping circles, each of diameter d, can be packed. These so called ``circle packing'' solutions consists of disks arranged on the vertices of a regular polygon \cite{mughal2}.  The ratio of the  diameter of the enclosing circle to the diameter of the disks is given by, 
\begin{equation}
\
D^c(N)/d
=
\left\{
\begin{array}{l l}
1 & \quad \mbox{if $N=1$}\\
 \left(1+\frac{1}{\sin\left(\pi/N \right) } \right)  
 & \quad \mbox{if $ N \geq 2$}, \\
\end{array}
\right.
\label{eq:Dvalues}
\end{equation}
where $D^c(2)/d=2$, $D^c(3)/d=2.1547$, $D^c(4)/d=2.4142$, $D^c(5)/d=2.7013$. Replacing the disks with spheres we are able to define a basic unit cell containing N spheres, the length of the cell is found to be  \cite{mughal2}, 
\begin{equation}
L^{c}(N)
=
\left\{
\begin{array}{l l}
d & \quad \mbox{if $N=1$}\\
\frac{d}{2}\sqrt{3 - \frac{(1-\cos(\pi/N))}{(1+cos(\pi/N))} } & \quad \mbox{if $ N \geq 2$}, \\
\end{array}
\right.
\nonumber
\end{equation}
and the twist angle is $\alpha^{c}(N)=\pi/N$. Each successive layer in the full structure can be generated by translating the spheres in the previous layer a distance $L^c(N)$ along the cylindrical axis and rotating them by an angle $\alpha^{c}(N)$. The case for N=3 (or packing 10) is illustrated in Figure~\ref{packing_images}. The screw periodic nature of the structure is highlighted by colouring spheres in the same unit cell  with the same colour. All such simple arrangements based on the circle packing solutions yield achiral structures.

By numerical means we have discovered further maximal contact packings that do not correspond to  the circle packing solutions. In most cases these are chiral and consist of unit cell containing a single sphere (e.g. structure 7 as shown in Figure~\ref{packing_images}). An exception to this is any maximal contact packing with phyllotactic indices $(l,l/2, l/2)$. Such packings are achiral and the only single-layer packing of this type is structure 13 (4,2,2), which is found at $D/d = 2.2247$ and has a unit cell with two spheres.

\begin{figure}
\begin{center}
\resizebox*{10cm}{!}{\includegraphics[angle=0]{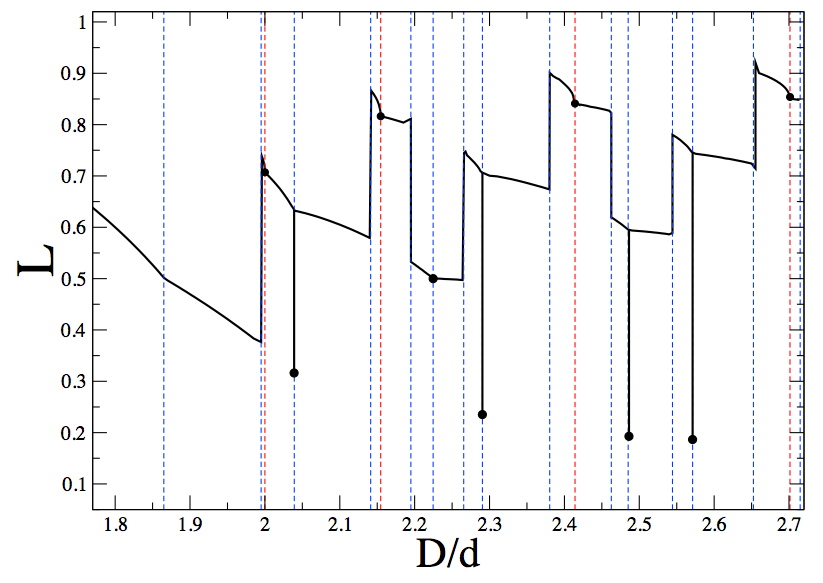}}
\resizebox*{10cm}{!}{\includegraphics[angle=0]{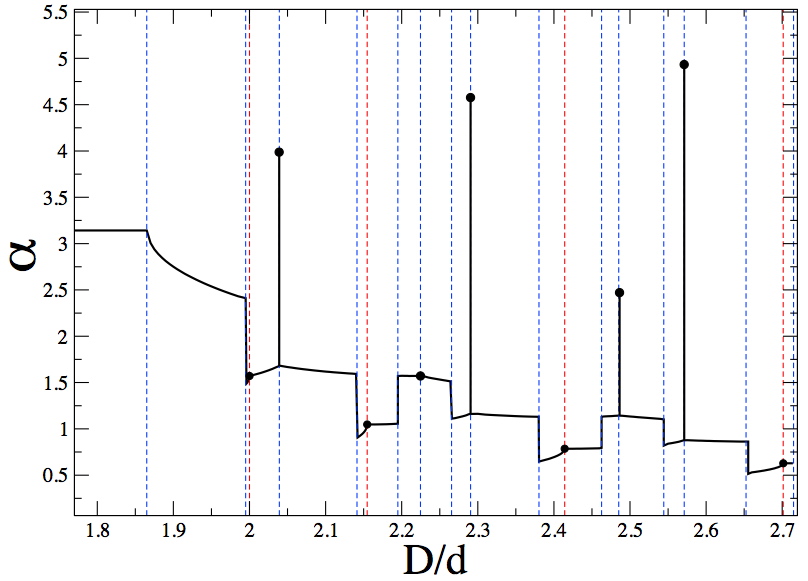}}
\resizebox*{10cm}{!}{\includegraphics[angle=0]{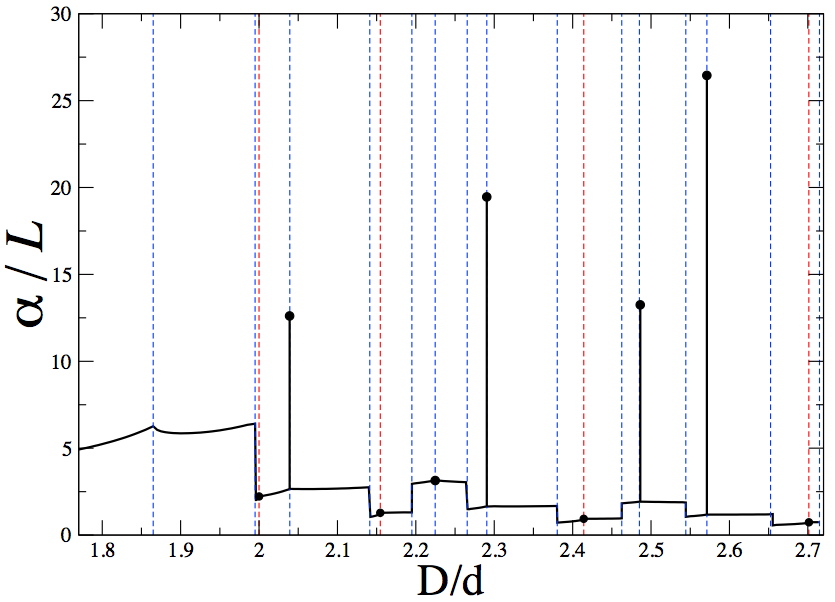}}
\end{center}
\caption{{\bf Top}:  $L$ plotted as a function of $D/d$. {\bf Middle}: $\alpha$ plotted as a function of $D/d$. {\bf Bottom}: the ratio of the previous two quantities, or twist per unit length, plotted as a function of $D/d$. The black line corresponds to the value of the relevant quantities for the various line slip structures while the black dots indicate the values for maximal contact packings. In some cases this involves an abrupt change in these values as explained in the text.}
\label{screw_symm}
\end{figure}

The values of the cell length $L_{max}$, twist $\alpha_{max}$ and twist per unit length $(\alpha/L)_{max}$ for all maximal contact monolayer packings are given by black dots in Figure~\ref{screw_symm}. The diameter ratio of the various circle-packing solutions, as given by  Eq. ($\!\!$~\ref{eq:Dvalues}), are indicated by red lines.

\subsection{Line-slip packings}

For values of $D/d$ which do not correspond to one of the maximal contact structures the optimal packing is of the line-slip type \cite{mughal1, mughal2}. Compared with maximal contact packings these have fewer number of contacts per sphere and involve a relative displacement between neighbouring spiral chains, which can be seen most directly in the rolled-out phyllotactic diagram.

The values of $L_{line}$, $\alpha_{line}$ and $(\alpha/L)_{line}$ for each of the line-slip packings are plotted in Figure~\ref{screw_symm} and are given by the black curve. Discontinuities in the first derivative of of these values indicate transitions between different line-slip structures.

Furthermore, at particular values of $D/d$, corresponding to particular maximal contact packings, the quantities plotted in Figure~\ref{screw_symm} are seen to jump suddenly in value. To understand this consider structure 6 which is the densest packing found in the range $2.0 < D/d \leq 2.039$. This is a line-slip arrangement that consists of a basic unit cell with two spheres. The full structure can be generated by means of the appropriate screw operation. This is illustrated in Figure~\ref{packing_images} and once again spheres in the same unit cell are given the same colour. 

The relative slip between neighbouring spiral chains in packing 6 proceeds until the next nearest neighbouring spheres are brought into contact. The result at this point is the chiral maximal contact packing 7 (at $D/d=2.039$), which compared to packing 6 has a higher degree of symmetry and has a unit cell containing only a single sphere.

In this manner, when a line-slip structure can be continuously deformed into a {\it chiral} maximal contact packing there is an discontinuous change in the cell length. The screw quantities are found to related as follows \cite{mughal3},
\begin{equation}
L_{line}=mL_{max},
\label{eq:mrule}
\end{equation}
 and 
 \[
  \alpha_{line}=(m\alpha_{max}) \; \textrm{mod} \; (2\pi),
 \]
 where $\textrm{mod}\; (2\pi)$ is the modulo operator with divisor $2\pi$ and $m$ is the second phyllotactic index. The second index is needed because all cases of this kind, so far observed, (whereby a line slip terminates in a chiral maximal contact packing) involve a line-slip of the type $(l,{\bf m}, n)$ \cite{mughal2}. In addition to giving the direction of the line slip the highlighted index also gives the number of sphere centres in the basic unit cell \cite{mughal2, mughal3}. These, as we have already pointed out, are reduced from $m$ to 1 when the next nearest neighbouring spheres are brought into contact and thus the relevant quantities are scaled by this factor.

\section{Helicity}

Recently, twisted periodic boundary conditions have also been used to investigate the ground state configuration of screened charges (interacting via a Yukawa potential) in a cylindrical tube \cite{oguz}. In the Yukawa study minimal energy arrangements are classified as being either helical or non-helical. The concept of helicity is widely used in biology and chemistry and here we briefly outline its relevance to columnar crystals.

Non-helical arrangements involve a unit cell with vanishing twist angle \cite{oguz}. In this sense any packing that can be decomposed into a unit cell with a twist angle that is a rational fraction of $2\pi$ is non-helical. In such cases it is possible to repeat the structure $X$ times to yield an arrangement with a twist that is a multiple of $2\pi$ (i.e. a structure with vanishing torsion). Thus if $\alpha=(2\pi)p/q$, where $p$ and $q$ are any positive integer, then we require,
\begin{eqnarray}
X\alpha&=&2\pi s,
\nonumber
\\
2\pi X\frac{p}{q}&=&2\pi s,
\nonumber
\end{eqnarray}
where $s$ is any positive integer. This can be solved to give,
\[
X=\frac{sq}{p},
\]
setting $s=p$ we have $X=q$. 

So for example, if $p=5$ and $q=8$ we find that 8 unit cells with, 
\[
\alpha=2\pi\frac{p}{q}=2\pi\frac{5}{8} 
\]
can be combined to produce a longer packing with a vanishing twist, since
\[ 
(2\pi s)\textrm{mod} \; (2\pi)=(10\pi)\textrm{mod} \; (2\pi)=0.  
\]

It can be immediately seen that all the achiral maximal contact packings (i.e. the circle packing solutions and structure 13) are non-helical. Helical arrangements, on the other hand, involve a irrational twist and and for these it is not possible to define a unit cell with simpler boundary conditions. Maximal contact chiral arrangements are of this type as are line-slip packings in general (although for a particular value of $D/d$ it may be that a line slip packing has a twist which by chance is a rational fraction of $2\pi$).

\section{Conclusion}
In summary we investigated the dense packing of hard spheres in an infinite cylindrical channel. We showed that such packings are comprised of a basic unit cell which can be repeated along the length of the tube by a screw displacement. We find that $\alpha$, $L$ and $\alpha/L$ are piecewise continuous function of the diameter ratio $D/d$, discontinuities are due transitions between various line-slip packings. Sudden jumps in the screw quantities are related to particular values of $D/d$ at which a complex packing (i.e. line slip arrangement with many spheres per unit cell) is coincident (or identical) with a much simpler packing (i.e. a chiral maximal contact packing with a single sphere in the unit cell). 

We showed that where the unit cell has a twist angle that is a rational fraction of $2\pi$ the cells can always be stacked together to form a longer packing with a vanishing torsion, such structures are by definition non-helical and include all the achiral maximal contact packings. Where this is not possible the packing can be defined as being helical and this includes all maximal contact chiral packing and line-slip arrangements (in general).

In this paper we have provided a report of some of our numerical results from simulations along with a sketch of the relevant analytical rules. Our future work will be to give these observations (including Eq. ($\!\!$~\ref{eq:mrule})) a firm theoretical foundation. 

\section{Acknowledgements}

AM acknowledges numerous useful discussions with Denis Weaire.


\label{lastpage}

\end{document}